\documentclass[runningheads]{llncs}
\usepackage{cite}
\usepackage{amsmath,amssymb,amsfonts}
\usepackage{algorithmic}
\usepackage{graphicx}
\usepackage{textcomp}
\usepackage{multirow}
\usepackage{array}
\usepackage{cite}
\usepackage{graphicx}
\usepackage{textcomp}
\usepackage{multirow}
\usepackage{array}
\usepackage{tikz}
\usepackage{footnote}
\usepackage[ruled,vlined, linesnumbered]{algorithm2e}
\usepackage[super]{nth}
\begin{document}
\title{DAL: Feature Learning from Overt Speech to Decode Imagined Speech-based EEG Signals with Convolutional Autoencoder\thanks{This research was supported by the Defense Challengeable Future Technology Program of Agency for Defense Development, Republic of Korea.}}
\titlerunning{Decoding Imagined Speech-based EEG Signals applying Feature Learning}
%
\author{Dae-Hyeok Lee\inst{1}\orcidID{0000-0002-2238-8910} \and
Sung-Jin Kim\inst{2}\orcidID{0000-0002-7918-3502} \and
Seong-Whan Lee\inst{2}\orcidID{0000-0002-6249-4996}}
\authorrunning{D.-H. Lee et al.}
%
\institute{Department of Brain and Cognitive Engineering, Korea University, Anam-dong, Seongbuk-ku, Seoul 02841, Korea \\
\email{lee\_dh@korea.ac.kr}
\and
Department of Artificial Intelligence, Korea University, Anam-dong, Seongbuk-ku, Seoul 02841, Korea \\
\email{s\_j\_kim@korea.ac.kr and sw.lee@korea.ac.kr}}

\maketitle              
\begin{abstract}
Brain-computer interface (BCI) is one of the tools which enables the communication between humans and devices by reflecting humans’ intention and status. With the development of artificial intelligence, the interest in communication between humans and drones using electroencephalogram (EEG) is increased. Especially, in the case of controlling drone swarms such as direction or formation, there are many advantages compared with controlling a drone unit. Imagined speech is one of the endogenous BCI paradigms, which can identify users’ intentions. When conducting imagined speech, the users imagine the pronunciation as if actually speaking. In contrast, overt speech is a task in which the users directly pronounce the words. When controlling drone swarms using imagined speech, complex commands can be delivered more intuitively, but decoding performance is lower than that of other endogenous BCI paradigms. We proposed the Deep-autoleaner (DAL) to learn EEG features of overt speech for imagined speech-based EEG signals classification. To the best of our knowledge, this study is the first attempt to use EEG features of overt speech to decode imagined speech-based EEG signals with an autoencoder. A total of eight subjects participated in the experiment. When classifying four words, the average accuracy of the DAL was 48.41\%. In addition, when comparing the performance between w/o and w/ EEG features of overt speech, there was a performance improvement of 7.42\% when including EEG features of overt speech. Hence, we demonstrated that EEG features of overt speech could improve the decoding performance of imagined speech.\\

\keywords{Deep autoencoder \and Drone swarms control \and Imagined speech.}
\end{abstract}
\section{INTRODUCTION}
Artificial intelligence (AI) has recently developed, and based on this, it has been widely used in various domains. In particular, advanced machine learning and pattern recognition methods have been used in a lot of research fields in the brain-computer interface (BCI) domain. BCI is one of the tools which enables the communication between humans and devices by reflecting humans’ intention and status~\cite{Suk2011subject, Lee2019eeg, Zhang2019making, Lee2020continuous, Wolpaw2002brain, Kwon2020subject, Lee2018high}. Recently, non-invasive BCI systems have been used for controlling external devices~\cite{Stawicki2017novel, Jeong2019trajectory, Meng2016noninvasive}. 

Advanced AI is important for the \nth{4} industrial revolution. In addition, interest in drone control is increasing due to the \nth{4} industrial revolution. The previous studies related to drone control were a camera-based vision approach or an approach using a joystick. Kaufmann \textit{et al}.~\cite{Kaufmann2018deep} developed the drone control system combining the convolutional neural networks with a path-planning and control system using camera-based vision data. Sanders \textit{et al}.~\cite{Sanders2018traditional} investigated the performance difference between the joystick- and gesture-based methods and indicated that the joystick-based method generally performs better than the other method. These approaches enable drone control with high accuracy, but when an unexpected situation occurs suddenly, there is a high probability of causing a serious error. In addition, there is a disadvantage that the flexibility according to the situation is lowered.

Accordingly, as a research field in the BCI domain, drone control using electroencephalogram (EEG) is one of the challenging research topics. Since EEG signals can reflect the user's intentions and current status, using EEG signals to control the drone has the advantage of being able to flexibly cope with unexpected situations. Lee \textit{et al}.~\cite{Lee2021design} designed endogenous BCI paradigms (motor imagery, visual imagery, and speech imagery) for acquiring EEG signals. EEG-based various tasks related to controlling drone swarm were instructed to subjects. Karavas \textit{et al}.~\cite{Karavas2017hybrid} showed the preliminary results of a hybrid brain-machine interface that combined information from an external device and the brain. They instructed the subjects to spread-out and fall-in drone swarms consisting of three drones.

BCI is largely divided into stimulus-based exogenous BCI and imagination-based endogenous BCI. Stimulus-based exogenous BCI has high accuracy but has a disadvantage in that the fatigue level of the body, especially the eyes, is quite high, and additional equipment that can give stimulation is required. On the other hand, if the imagination-based endogenous BCI is used, the disadvantages of exogenous BCI can be solved. Imagined speech is one of the endogenous BCI paradigms, which can identify users’ intentions. When conducting imagined speech, the users imagine the pronunciation as if actually speaking. Bakhshali \textit{et al}.~\cite{Bakhshali2020eeg} analyzed the various metrics and methods of effectively extracting EEG features of imagined speech and they show that Riemannian distance effectively extracts features in EEG signals. Nguyen \textit{et al}.~\cite{Nguyen2017inferring} analyzed that different characteristics exist depending on the characteristics of the word they imagine by mapping EEG signals into the Riemannian manifold. In addition, overt speech is a task in which the users directly pronounce the words. Watanabe \textit{et al}. ~\cite{Watanabe2020synchronization} analyzed the similarity between overt speech- and imagined speech-based EEG signals using the statistical methods. Sereshkeh \textit{et al}.~\cite{Sereshkeh2017online} proposed the model using a support vector machine (SVM) and a combination of spectral and time-frequency features to decode overt speech-based EEG signals and got 75.9\% accuracy at binary classification. However, many studies related to decoding imagined speech-based EEG signals usually show low performances. In addition, most of the studies have been analyzed using machine learning algorithms rather than deep learning-based models. Lee \textit{et al}.~\cite{Lee2019towards} analyzed imagined speech-based EEG signals in the perspective of its presence, spatial features. They conducted the classification using the shrinkage regularized linear discriminant analysis and the random forest (RF) which are general machine learning algorithms. Hern{\'a}ndez-Del-Toro \textit{et al}.~\cite{Hernadez2021toward} decoded imagined speech-based EEG signals using five feature extraction methods to solve the task of detecting imagined words segments from continuous EEG signals. They tested in three datasets using four machine learning algorithms (RF, k-nearest neighbors, SVM, and logistic regression).

In this paper, we propose the framework, called Deep-autolearner (DAL) to increase the performance of decoding imagined speech-based EEG signals. Our proposed framework could learn EEG features of overt speech for imagined speech-based EEG signals classification by using the architecture including the encoder and decoder. By designing the architecture using the encoder and decoder, the decoder could assist to train the encoder to extract significant EEG features of imagined speech which have a high correlation with overt speech-based EEG signals. To the best of our knowledge, this study is the first attempt to use the autoencoder by applying the important EEG features of overt speech to extract EEG features of imagined speech. The proposed framework has achieved the best performance (48.41\% ($\pm$6.16)). In addition, when comparing the performance between DAL using imagined speech-based EEG signals only (w/o overt speech) and DAL including EEG features of overt speech (w/ overt speech) when training the model, there is a performance improvement of 7.42\% in the case of w/ overt speech. Hence, we demonstrated that EEG features of overt speech could improve the decoding performance of imagined speech.

The rest of this paper is organized as follows. In Section 2, we introduce our experimental design and the structure of the proposed model more in detail. In Section 3, we present the experimental results. In Section 4, we present the directions for decoding imagined speech-based EEG signals and the limitations of our study. Finally, we conclude this paper in Section 5.\\
\section{MATERIALS AND METHODS}
\subsection{Subjects}
A total of eight healthy subjects (S1--S8, six males and two females, aged 26.2 ($\pm 2.7$)) participated in our experiment. The Institutional Review Board at Korea University (KUIRB-2020-0318-01) reviewed and approved the experimental environment and protocols. 
\begin{figure}[t!]
\includegraphics[width=\columnwidth]{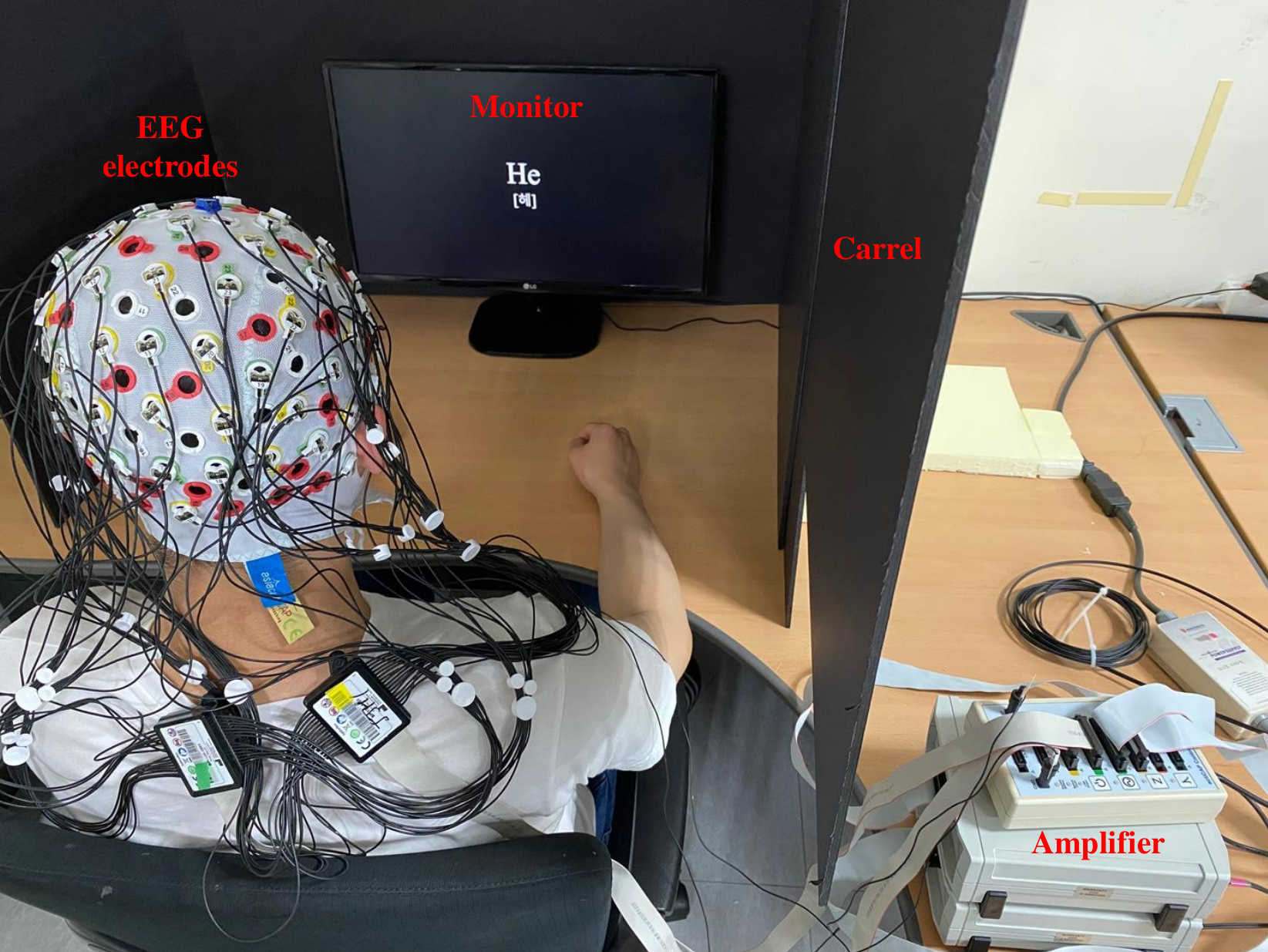}
\caption{Experimental environment for acquiring overt and imagined speech-based EEG signals. Installation of a black partition in the experimental space was for the subjects' concentration.} \label{fig1}
\end{figure}
Before the experiment, we informed all subjects of the experimental protocols, and after that, they consented according to the Declaration of Helsinki. In addition, the subjects were informed about getting adequate sleep (over seven hr.) and avoiding any alcohol the day before the experiment.\\
\subsection{Experimental Environment}
We used the signal amplifier (BrainAmp, Brain Products GmbH, Germany) for measuring subjects' EEG signals. We set up the sampling frequency to 1,000 Hz and a 60 Hz notch filter was applied for removing DC noise. We used 58 EEG channels for acquiring EEG signals that were placed on the subjects' scalp according to the international 10/20 system. In addition, we measured EOG signals by attaching six electrodes around the subjects' eyes. The FCz and FPz channels were used for reference and ground electrodes, respectively. The impedance of all EEG electrodes was set to lower 10 k$\Omega$ or less by injecting the conductive gel into the subjects' scalp before acquiring EEG signals.\\
\begin{figure}[t!]
\includegraphics[width=\textwidth]{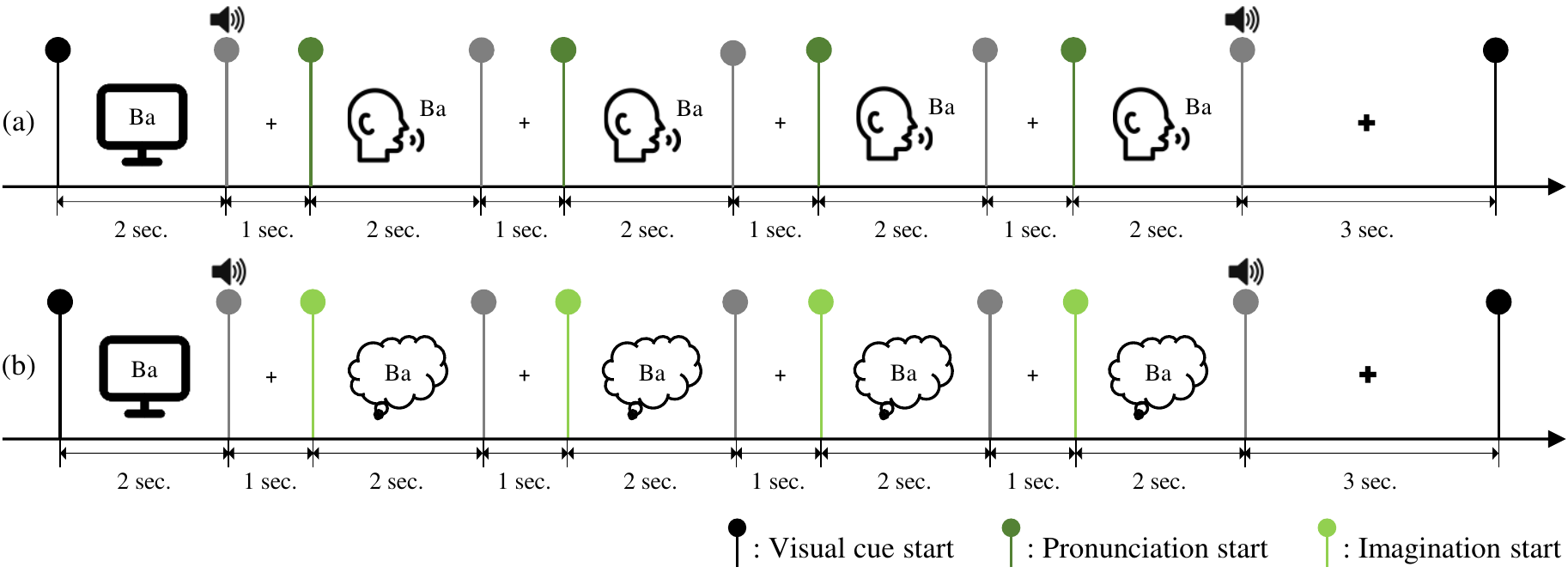}
\caption{Experimental paradigms for acquiring overt and imagined speech-based EEG signals. (a) The paradigm for acquiring overt speech-based EEG signals. (b) The paradigm for acquiring imagined speech-based EEG signals.} \label{fig2}
\end{figure}
\subsection{Experimental Paradigm}
We designed the experimental paradigms to acquire overt and imagined speech-based EEG signals of good quality. We selected four words (`Ba', `Ku', `He', and `Li') applying Levenshtein distance and the Soundex algorithm. The selection of words with a large difference in each index value is important, so we selected four words based on these criteria. Three phases existed in the experiment. The experiments of overt and imagined speech were the same process. We instructed the subjects to perform overt speech experiment first and then imagined speech experiment. One of four words was displayed over 2 sec. randomly. A fixation cross was provided for 1 sec. and a blank image was represented for 2 sec.. We instructed the subjects to perform a task (pronunciation or imagination) when a blank image was displayed. A fixation cross and a blank image were repeated four times per word. After that, a bold fixation cross appeared for 3 sec. to eliminate the feel of the existing word. We acquired 50 trials per word (a total of 200 trials).

To control the drone swarm with high degrees of freedom using imagined speech, using acoustic-based imagined speech is an essential part of drone swarm control. For this reason, by using this paradigm which includes words presenting the significant features of acoustic, we investigated the decoding method based on imagined speech.\\
\subsection{Deep-autolearner (DAL)}
Fig. 3 shows an overview of the proposed model for the classification of imagined speech-based EEG signals. The proposed framework extracts the feature vector of imagined speech from EEG signals through the encoder and then reconstructs the signals of overt speech using extracted feature vector at the decoder. Through the above process, the model could learn the significant common features of EEG signals between overt speech and imagined speech to classify imagined speech-based EEG signals.

Imagined speech-based EEG signals are entered into the encoder for extracting a feature vector, and this feature vector uses as the input format of the classifier and decoder. At the decoder, the model reconstructs EEG signals using the extracted feature vector to learn EEG features correlated with overt speech. When the decoder reconstructs the signals, concatenate boxes (C-Box) are used to deliver the information of EEG signals from imagined speech.
\begin{figure*}[t!]
\includegraphics[width=\textwidth]{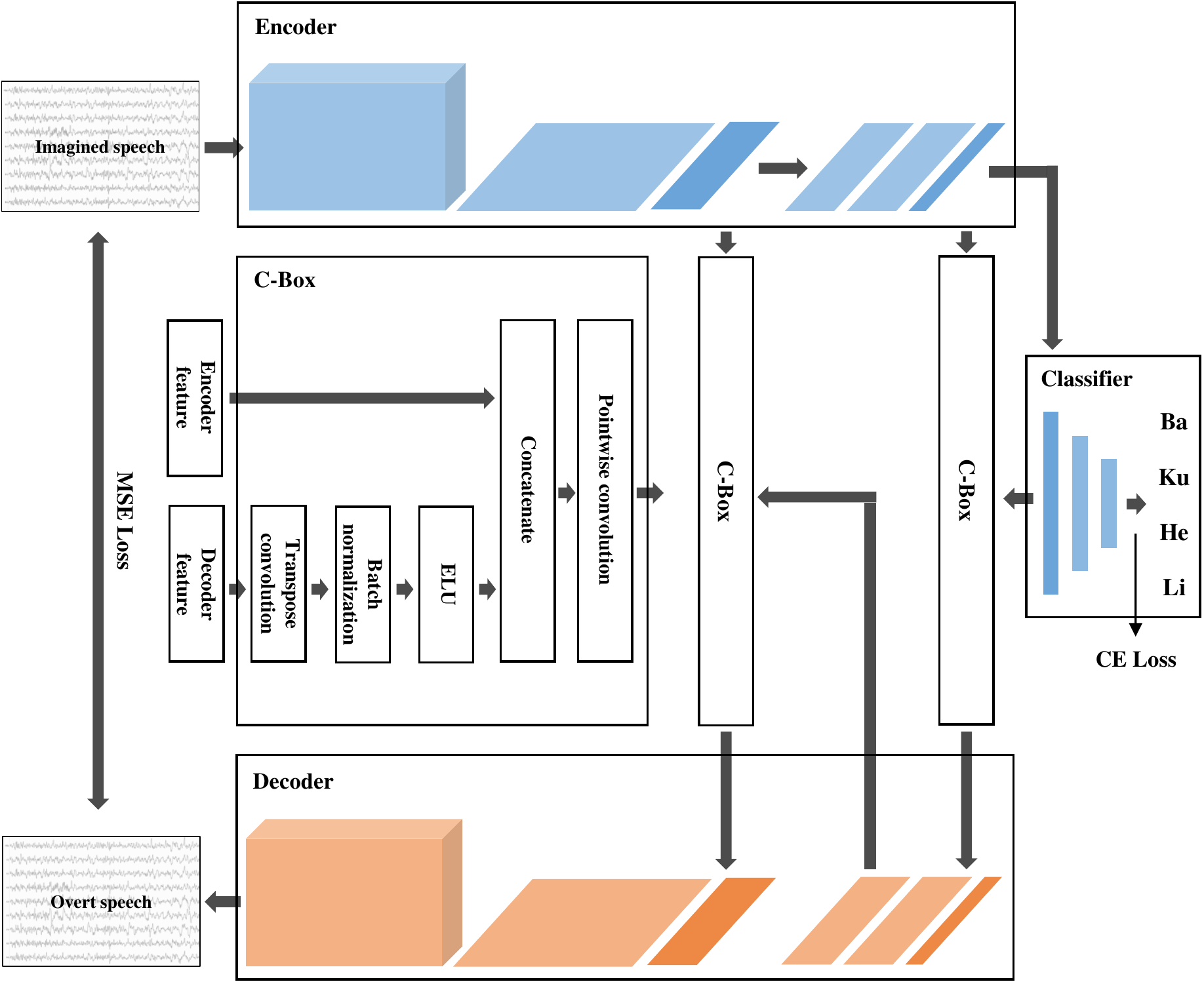}
\caption{The architecture of the proposed framework (DAL). EEG signals of imagined speech are used as the input representation of the DAL. The feature vector of imagined speech is of the encoder, and it is used as the input value of the classifier and decoder. Next, by using C-Box between the encoder and decoder, the decoder regenerates overt speech-based EEG signals from the extracted feature vector without losing the information of imagined speech.} \label{fig3}
\end{figure*}
When extracting features from EEG signals, raw signal is commonly used as input representation~\cite{Schirrmeister2017deep}. For this reason, as shown in the left part of Fig. 4, the encoder is developed based on the EEGNet~\cite{Lawhern2018eegnet}, which uses raw EEG signals as input and shows robust performance in decoding EEG signals. Also, similar to the EEGNet, the encoder consists of two blocks. At the first block, two convolution layers are used to extract temporal and spatial features. 
\begin{figure*}[t!]
\includegraphics[width=\textwidth]{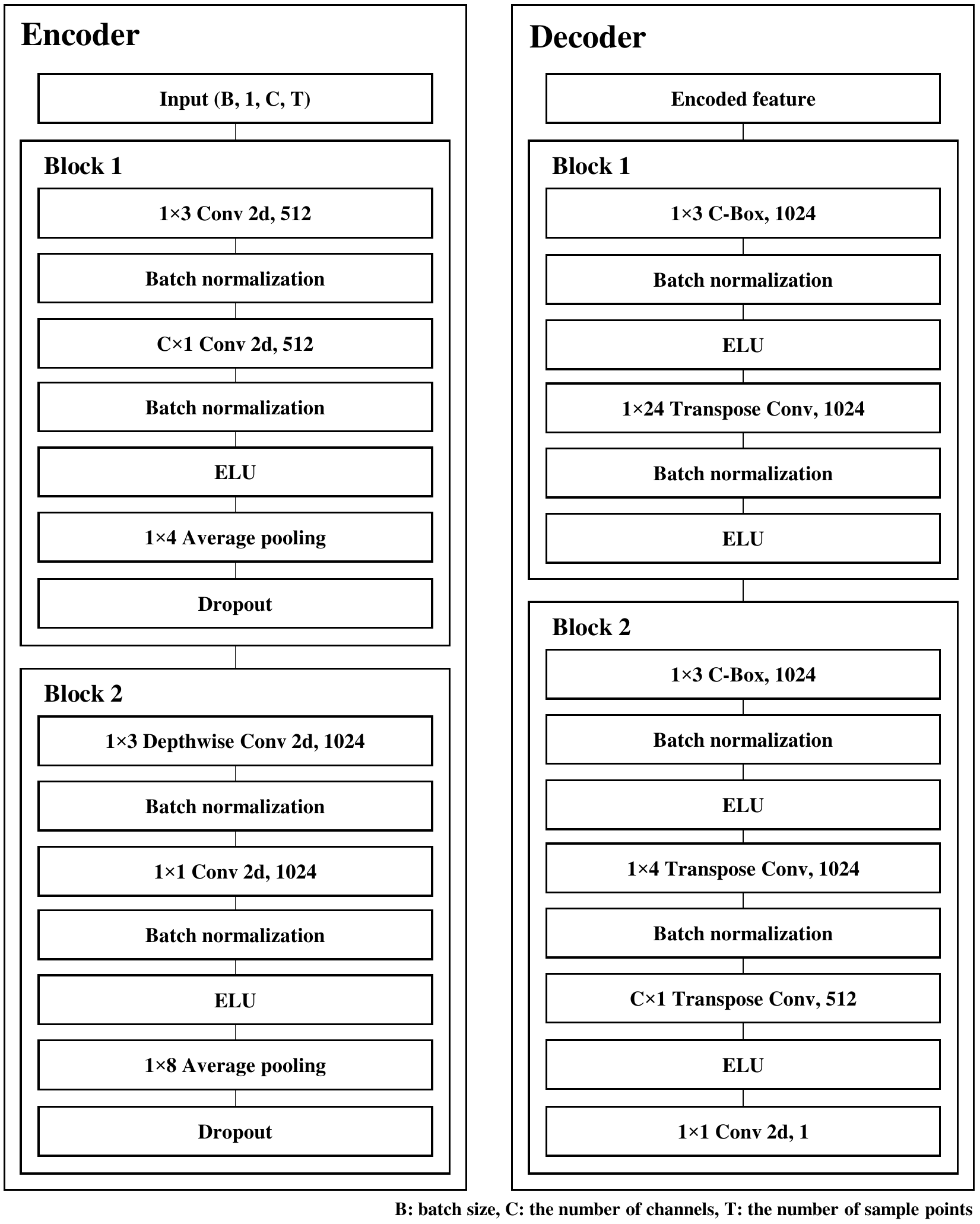}
\caption{Details of encoder and decoder. At the encoder, convolution layers with various kernel sizes extract different EEG features of imagined speech. Also, the decoder reconstructs temporal features of overt speech at first. After reconstructing temporal features, spatial features are regenerated.} \label{fig4}
\end{figure*}
The convolution layer can learn certain features depending on the size of the kernel. To get temporal features by channel, the filter size is set to one-three (1$\times$3) which only observes temporal features in one channel. After extracting temporal features, temporal properties are mixed channel-wise using spatial filters of size 58$\times$1. In the second block, depthwise convolution and pointwise convolution layers are used for enhancing discriminative features. Depthwise convolution can learn features by the depth and pointwise convolution can effectively fuse features in the depth direction. All blocks use exponential linear units (ELUs) and average pooling to improve nonlinearities and get summarized features. Also, to obtain generalized features, batch normalization (BN) and dropout are used.

The autoencoder is the neural network that learns useful representations by encoding and regenerating the input data~\cite{Schmidhuber2015deep}. Motivated by the concept of autoencoder, as shown in the right part of Fig. 4, the decoder of the proposed model is constructed to learn features of EEG signals of overt speech by reconstructing EEG signals from the feature vector of imagined speech. At the first block of the decoder, the feature vector of imagined speech is used as input representation, and the transpose convolution layer using filters of size 1$\times$24 is used to regenerate the temporal information of overt speech. To reconstruct EEG signals without losing information of imagined speech, we proposed C-Box which is motivated by U-Net~\cite{Ronneberger2015unet}. At the C-Box, both features of encoder and decoder are used as input components, and the transpose convolution layer is used to make the shape of features equal before concatenating both features. By attaching the C-Box between encoder and decoder, we can effectively rebuild EEG signals without loss of information. Next, at the second block of decoder, the transpose convolution layer with filters of size 1$\times$4 makes the model regenerate temporal information more detail, and the transpose convolution layer with filters of size 58$\times$1 reconstructs the spatial information. After two kinds of the transpose convolution layers, the pointwise convolution fuses features in the depth direction to get overt speech-based EEG signals. We also used C-Box to concatenate both features of the encoder and decoder. Also, ELUs and BN are used the same as the encoder. By conducting this process, the model could learn the significant common features of EEG signals between imagined speech and overt speech.
\begin{equation}
\mathcal{L}_{CE} = -\frac{1}{n}\sum_{i=1}^{n}y_{i}^{class}logC(E(x_{i}))
\end{equation}
\begin{equation}
\mathcal{L}_{MSE} = \frac{1}{n}\sum_{i=1}^{n}|y_{i}^{overt}-D(E(x_{i}))|
\end{equation}
\begin{equation}
\mathcal{L}_{total} = \alpha\mathcal{L}_{CE} + (1-\alpha)\mathcal{L}_{MSE}
\end{equation}

Our proposed model has two parts of the output. The first is the classification part, and the other is the generation part. To train the classification part and generation part simultaneously, we designed the loss function. Cross entropy (CE) is one of the most commonly used loss functions at the classification task. CE is the loss function based on information theory, and it calculates the difference of entropy between the predicted value and ground truth. Also, mean square error (MSE) is the loss function for regression that calculates the difference between the predicted value and ground truth based on the L1-norm. For using both properties of loss function at once, we combine both loss values. We used the value which applies the encoder $E$ and the classifier $C$ to the $i^{th}$ input data $x_{i}$ as the predicted classification value of the CE part and used $y_{i}^{class}$ represented with a one-hot vector corresponding to $x_{i}$ as the label vector (1). In addition, generated overt speech based-EEG signals applying the encoder $E$ and the decoder $D$ to input data $x_{i}$ are put into the MSE part and the difference from the real EEG signals of overt speech $y_{i}^{overt}$ is calculated (2). After getting both CE loss $\mathcal{L}_{CE}$ and MSE loss $\mathcal{L}_{MSE}$, we combine these loss values at a constant rate $\alpha$ to establish an appropriate learning ratio between the two values (3). In this paper,  we set $\alpha$ to 0.9. By designing loss function as shown in (3), we could train our model to learn only proper overt speech features needed for the classification of imagined speech data without bias on such classification or regression tasks.\\

\section{EXPERIMENTAL RESULTS}
\begin{table}[t!]
\caption{Comparison of classification accuracies for decoding imagined speech-based EEG signals using various methods}
\centering
\tiny
\renewcommand{\arraystretch}{1.7}
\resizebox{\columnwidth}{!}{
\begin{tabular}{c|cc|cc|cc}
\hline
\multirow{3}{*}{Subject} & \multicolumn{2}{c|}{CSP-LDA~\cite{Cho2018classification}}                                                                                           & \multicolumn{2}{c|}{EEGNet~\cite{Lawhern2018eegnet}}                                                                                            & \multicolumn{2}{c}{Proposed}                                                                                           \\ \cline{2-7} 
                         & \begin{tabular}[c]{@{}c@{}}w/o overt\\ speech\end{tabular} & \begin{tabular}[c]{@{}c@{}}w/ overt\\ speech\end{tabular} & \begin{tabular}[c]{@{}c@{}}w/o overt\\ speech\end{tabular} & \begin{tabular}[c]{@{}c@{}}w/ overt\\ speech\end{tabular} & \begin{tabular}[c]{@{}c@{}}w/o overt\\ speech\end{tabular} & \begin{tabular}[c]{@{}c@{}}w/ overt\\ speech\end{tabular} \\ \hline
S1                       & 32.43                                                      & 29.92                                                     & 32.94                                                      & 30.74                                                     & 37.06                                                      & 42.59                                                     \\
S2                       & 29.77                                                      & 25.66                                                     & 32.90                                                      & 34.04                                                     & 35.04                                                      & 41.22                                                     \\
S3                       & 33.72                                                      & 27.63                                                     & 33.05                                                      & 32.07                                                     & 36.72                                                      & 45.67                                                     \\
S4                       & 43.76                                                      & 33.24                                                     & 39.31                                                      & 38.82                                                     & 54.08                                                      & 58.83                                                     \\
S5                       & 23.51                                                      & 26.97                                                     & 33.55                                                      & 32.23                                                     & 39.33                                                      & 49.77                                                     \\
S6                       & 29.45                                                      & 23.69                                                     & 32.72                                                      & 33.38                                                     & 36.36                                                      & 46.16                                                     \\
S7                       & 23.86                                                      & 27.14                                                     & 35.21                                                      & 35.69                                                     & 39.33                                                      & 47.15                                                     \\
S8                       & 44.25                                                      & 33.89                                                     & 33.38                                                      & 32.24                                                     & 50.03                                                      & 55.86                                                     \\ \hline
Avg.                     & \textbf{32.59}                                             & \textbf{28.52}                                            & \textbf{34.13}                                             & \textbf{33.65}                                            & \textbf{40.99}                                             & \textbf{48.41}                                            \\
Std.                     & \textbf{7.91}                                              & \textbf{3.58}                                             & \textbf{2.24}                                              & \textbf{2.57}                                             & \textbf{7.06}                                              & \textbf{6.16}                                             \\ \hline
\end{tabular}}
\end{table}
\begin{figure*}[t!]
\includegraphics[width=\textwidth]{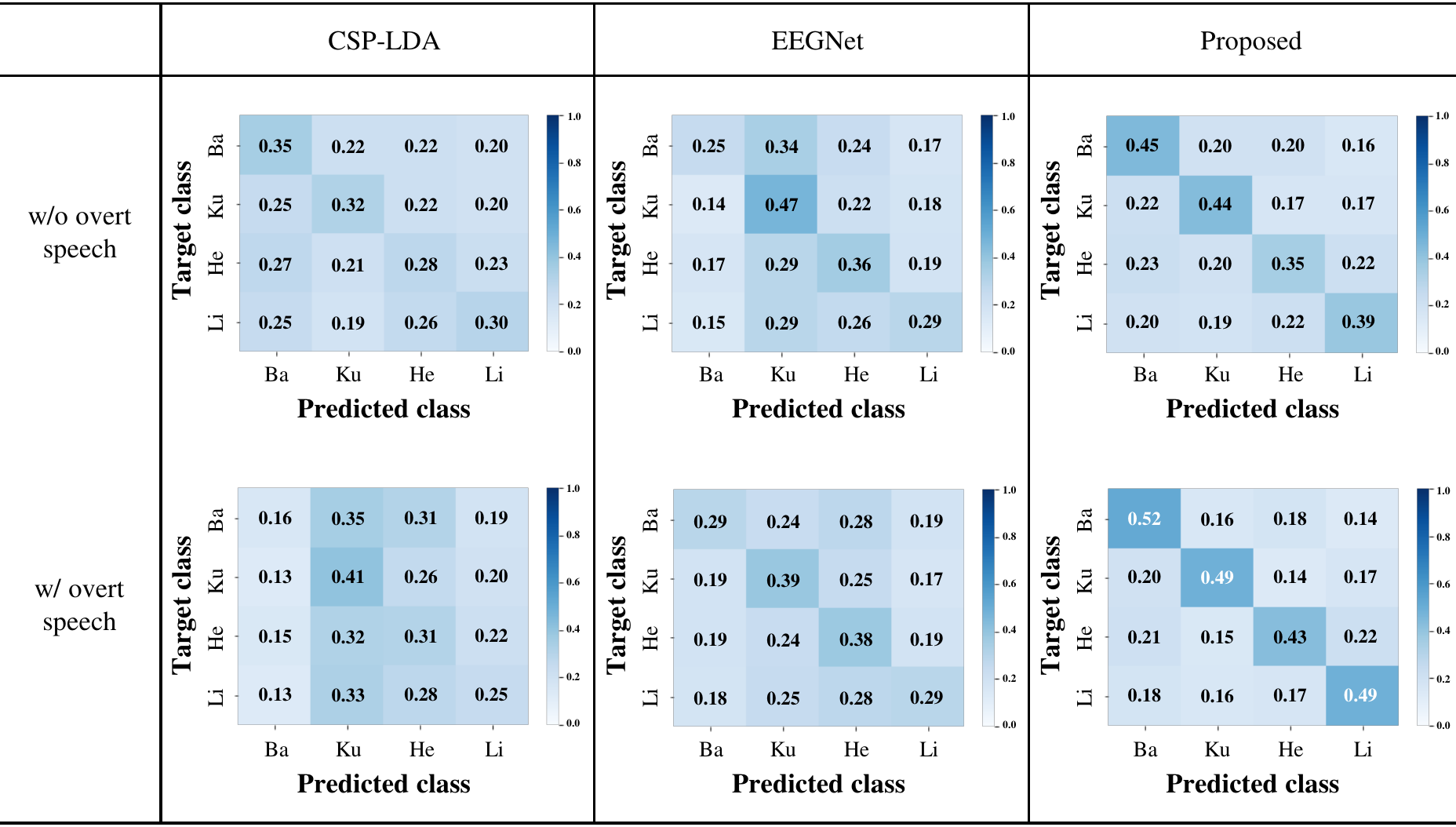}
\caption{Confusion matrices of classification performance for each word using the CSP-LDA w/o and w/ EEG features of overt speech, the EEGNet w/o and w/ EEG features of overt speech, and the DAL w/o and w/ EEG features of overt speech.} \label{fig5}
\end{figure*}
Table 1 represents the overall classification performances for each subject. We applied the five-fold cross-validation to evaluate classification accuracy fairly; 80\% of the whole samples were randomly selected for constructing the training set and the remaining 20\% were used for constructing the validation set. In addition, we repeated the five-fold cross-validation four times after adopting a different shuffle order each time ~\cite{Berrar2019cross}. We obtained the average accuracy of 40.99\% ($\pm$7.06) in the case of w/o overt speech. Also, the average accuracy of DAL w/ overt speech was 48.41\% ($\pm$6.16). S4 showed the highest accuracies of 54.08\% and 58.83\% in the case of DAL w/o and w/ overt speech, respectively. In contrast, S2 represented the lowest accuracies of 35.04\% and 41.22\% in the two cases, respectively. 

In addition, Table 1 shows the comparison of classification accuracies between the conventional methods and our framework. The conventional methods used for performance comparison were the common spatial pattern~\cite{Ang2008fbcsp}-linear discriminant analysis~\cite{Cho2018classification} (CSP-LDA) and the EEGNet~\cite{Lawhern2018eegnet}. The CSP-LDA extracted informative spatial features used the CSP, and the LDA was used as a classifier. The EEGNet is characterized by a small number of parameters, using depthwise separable convolution. The network consists of temporal convolution and spatial convolution blocks using depthwise convolution, followed by depthwise separable convolution blocks. When training the model w/ overt speech, the CSP-LDA and the EEGNet showed a different tendency from the DAL. The average accuracy of our proposed DAL improved when training the model w/ overt speech, but the average accuracies of the CSP-LDA and the EEGNet deteriorated. In other words, the average accuracy of the CSP-LDA w/o overt speech was higher than that of the CSP-LDA w/ overt speech, and the average accuracy of the EEGNet w/o overt speech was higher than that of the EEGNet w/ overt speech. In the case of our proposed DAL, S5 showed the largest accuracy improvement, and the value was 10.44\%. In contrast, for the CSP-LDA, S4 represented the largest accuracy decrease of 10.52\%, and in the case of the EEGNet, S1 showed the largest accuracy decrease of 2.20\%. In the case of S1, S2, S3, S4, S6, and S8 of the CSP-LDA and S1, S3, S4, S5, and S8 of the EEGNet, EEG features of overt speech deteriorated the accuracies when training the model. The average accuracies of the CSP-LDA w/o overt speech and w/ overt speech were 32.59\% ($\pm$7.91) and 28.52\% ($\pm$3.58), respectively. For the EEGNet w/o overt speech and w/ overt speech, the average accuracies were 34.13\% ($\pm$2.24) and 33.65\% ($\pm$2.57), respectively. Through these results, we showed that the average accuracies of our proposed model were the highest in both w/o and w/ overt speech cases compared with those of the conventional methods. 

We also evaluated the degree of confusion using all subjects. Fig. 5 shows the confusion matrices when classifying four words (`Ba', `Ku', `He', and `Li') using the CSP-LDA w/o and w/ EEG features of overt speech, the EEGNet w/o and w/ EEG features of overt speech, and the DAL w/o and w/ EEG features of overt speech. Each row of the matrix contains the target states and each column represents the predicted states. The DAL w/ EEG features of overt speech showed the highest true positive rate for classifying all words, and the values were 0.52, 0.49, 0.43, and 0.49, respectively. 
\begin{figure*}[t!]
\includegraphics[width=\textwidth]{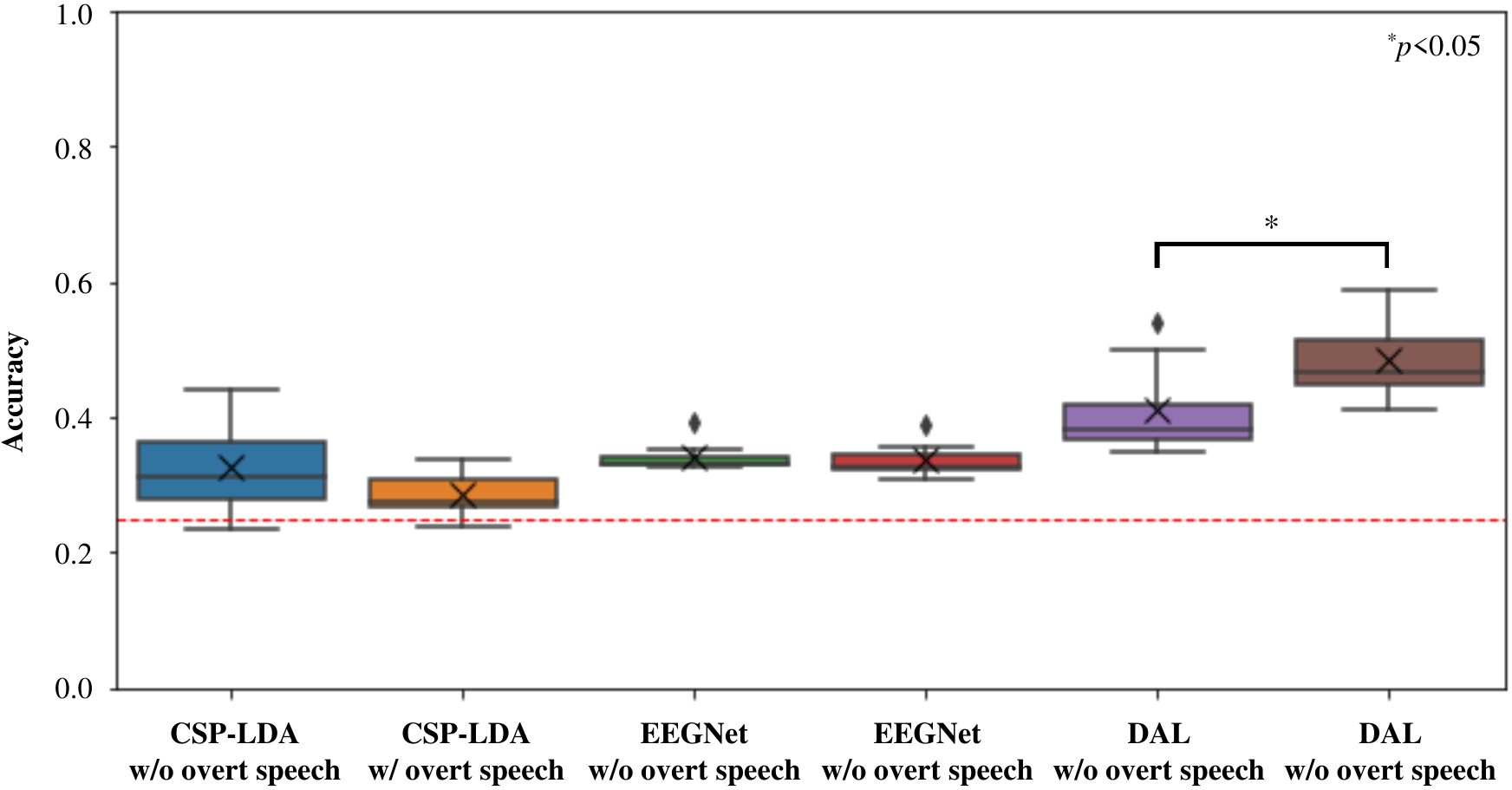}
\caption{Results of decoding normalized to the average performance across all decoding models including the statistical analyses. The red-colored dotted line indicates the chance-level accuracy of classifying 4-class.} \label{fig6}
\end{figure*}

To verify the difference of classification performance between each model w/o and w/ EEG features of overt speech, we applied the analysis of variance with the Bonferroni correction for multiple comparisons. Initially, we investigated normality and homoscedasticity due to a small number of samples. The normality for each method w/o and w/ EEG features of overt speech applying the Shapiro–Wilk test was satisfied with a null hypothesis. In addition, the assumption of homoscedasticity based on Levene’s test was also met for each group. After investigating the normality and homoscedasticity, we conducted a statistical analysis between each model w/o and w/ EEG features of overt speech which satisfied these conditions. We could confirm the statistically significant difference between the DAL w/o and w/ EEG features of overt speech (\textit{p}$<$0.05) as shown in Fig. 6. In contrast, no statistically significant difference existed between the CSP-LDA w/o and w/ EEG features of overt speech and the EEGNet w/o and w/ EEG features of overt speech (\textit{p}$>$0.05). 
\section{DISCUSSION}
This study shows the possibility to increase the decoding performance of imagined speech-based EEG signals by using overt speech-based EEG signals. To this end, we propose a framework that could learn EEG features of overt speech for imagined speech-based EEG signals classification by using the architecture including the encoder and decoder. The purpose of using a decoder is to assist in training the encoder for extracting significant EEG features of imagined speech which have a high correlation with overt speech-based EEG signals. Hence, we believe that this study could contribute to communication using imagined speech-based EEG signals.

We decoded imagined speech-based EEG signals w/o EEG features of overt speech and w/ EEG features using three models (CSP-LDA, EEGNet, and our proposed DAL). In each model w/o EEG features of overt speech, the DAL showed the highest performance of 40.99\%, followed by the EEGNet (34.13\%) and the CSP-LDA(32.59\%). This tendency was the same for each model w/ EEG features of overt speech (28.52\%, 33.65\%, and 48.41\%, respectively). Interestingly, regardless of whether EEG features of overt speech were not included, the standard deviation of the EEGNet was significantly lower than that of other models. In addition, in the case of the CSP-LDA, when EEG features of overt speech were included, the standard deviation dropped by 4.33, which was a fairly high value. Also, when training the model w/ overt speech, the CSP-LDA and the EEGNet showed a different tendency from the DAL. The average accuracy of the DAL improved when training the model w/ overt speech, but the average accuracies of the CSP-LDA and the EEGNet deteriorated. That means, our proposed framework is specialized for training EEG features of overt speech compared to other models.

However, there are still some issues that remain. We will continue to develop our proposed framework. First, the proposed framework has a relatively high standard deviation. In other words, this means that the stability of the model needs to be supplemented. Also, in order to utilize our framework in a real-world environment, higher decoding performance is required.\\
\section{CONCLUSION}
Decoding imagined speech-based EEG signals with high performance is one of the important challenging issues. The conventional related studies have decoded imagined speech-based EEG signals applying machine learning algorithms. In the BCI domain, data acquisition is hard compared to other domains. In the case of analyzing EEG signals using handcraft features-based machine learning algorithms, the information loss occurs with high probability. Hence, deep learning-based approaches that can analyze without loss of information are also required when decoding imagined speech-based EEG signals.

In this work, we present the deep learning-based framework to increase the performance of decoding imagined speech-based EEG signals. The proposed framework could learn EEG features of overt speech for decoding imagined speech-based EEG signals. To the best of our knowledge, this study is the first attempt to use EEG features of overt speech to decode imagined speech-based EEG signals with an autoencoder. Therefore, we demonstrated the feasibility of communication using imagined speech-based EEG signals.\\

%
%
%
\bibliographystyle{splncs04}

\begin{thebibliography}{30}
\bibitem{Suk2011subject}
Suk, H.-I. and Lee, S.-W.: Subject and class specific frequency bands selection for multiclass motor imagery classification. Int. J. Imaging Syst. Technol. 21(2), 123--130 (2011).

\bibitem{Lee2019eeg}
Lee, M.-H., Kwon, O.-Y., Kim, Y.-J., Kim, H.-K., Lee, Y.-E., Williamson, J., Fazli, S., and Lee, S.-W.: EEG Dataset and OpenBMI Toolbox for Three BCI Paradigms: An Investigation into BCI Illiteracy. GigaScience 8(5), 1--16 (2019).

\bibitem{Zhang2019making}
Zhang, D., Yao, L., Chen, K., Wang, S., Chang, X., and Liu, Y.: Making sense of spatio-temporal preserving representations for EEG-based human intention recognition. IEEE Trans. Cybern. 50(7), 3033--3044 (2019).

\bibitem{Lee2020continuous}
Lee, D.-H., Jeong, J.-H., Kim, K., Yu, B.-W., and Lee, S.-W.: Continuous EEG Decoding of Pilots’ Mental States Using Multiple Feature Block-Based Convolutional Neural Network. IEEE Access 8, 121929--121941 (2020).

\bibitem{Wolpaw2002brain}
Wolpaw, J.  R., Birbaumer, N., McFarland, D. J., Pfurtscheller, G., and Vaughan, T. M.: Brain-computer interfaces for communication and control. Clin. Neurophysiol. 113(6), 767--791 (2002).

\bibitem{Kwon2020subject}
Kwon, O.-Y., Lee, M.-H., Guan, C., and Lee, S.-W.: Subject-Independent Brain-Computer Interfaces Based on Deep Convolutional Neural Networks. IEEE Trans. Neural Netw. Learn. Syst. 31(10), 3839--3852 (2020).

\bibitem{Lee2018high}
Lee, M.-H., Williamson, J., Won, D.-O., Fazli, S., and Lee, S.-W.: A high performance spelling system based on EEG-EOG signals with visual feedback. IEEE Trans. Neural Syst. Rehabil. Eng. 26(7), 1443--1459 (2018).

\bibitem{Stawicki2017novel}
Stawicki, P., Gembler, F., Rezeika, A., and Volosyak, I.: A Novel Hybrid Mental Spelling Application Based on Eye Tracking and SSVEP-Based BCI. Brain Sci. 7(4), 35--51 (2017).

\bibitem{Jeong2019trajectory}
Jeong, J.-H., Shim, K.-H., Kim, D.-J., and Lee, S.-W.: Trajectory Decoding of Arm Reaching Movement Imageries for Brain-Controlled Robot Arm System. Annual Int. Conf. IEEE Eng. Med. and Bio. Soc. (EMBC), Germany, pp. 1--4 (2019).

\bibitem{Meng2016noninvasive}
Meng, J., Zhang, S., Bekyo, A., Olsoe, J., Baxter, B., and He, B.: Noninvasive Electroencephalogram Based Control of a Robotic Arm for Reach and Grasp Tasks. Sci. Rep. 6, 1--15 (2016). 

\bibitem{Kaufmann2018deep}
Kaufmann, E., Loquercio, A., Ranftl, R., Dosovitskiy, A., Koltun, V., and Scaramuzza, D.: Deep Drone Racing: Learning Agile Flight in Dynamic Environments. Conf. Robot Learn. (CoRL), Zurich, pp. 133--145 (2018).

\bibitem{Sanders2018traditional}
Sanders, B., Vincenzi, D., Holley, S., and Shen, Y.: Traditional Vs Gesture Based UAV Control. Adv. Intell. Syst. Comput. 784, 15--23 (2018).

\bibitem{Lee2021design}
Lee, D.-H., Jeong, J.-H., Ahn, H.-J., and Lee, S.-W.: Design of an EEG-based Drone Swarm Control System using Endogenous BCI Paradigms. Int. Winter Conf. Brain-Computer Interface (BCI), Korea, pp. 1--5 (2021).

\bibitem{Karavas2017hybrid}
Karavas, G. K., Larsson, D. T., and Artemiadis, P.: A hybrid BMI for control of robotic swarms: Preliminary results. IEEE/RSJ Int. Conf. Intell. Robot. Syst. (IROS), Canada, pp. 5065--5075 (2017).

\bibitem{Bakhshali2020eeg}
Bakhshali, M. A., Khademi, M., Ebrahimi-Moghadam, A., and Moghimi, S.: EEG signal classification of imagined speech based on Riemannian distance of correntropy spectral density. Biomed. Signal Process. Control 59, 101899 (2020).

\bibitem{Nguyen2017inferring}
Nguyen, C. H., Karavas, G. K., and Artemiadis, P.: Inferring imagined speech using EEG signals: a new approach using Riemannian manifold features. J. Neural Eng. 15(1), 1--16 (2017).

\bibitem{Watanabe2020synchronization}
Watanabe, H., Tanaka, H., Sakti, S., and Nakamura, S.: Synchronization between overt speech envelope and EEG oscillations during imagined speech. Neurosci. Res. 153, 48--55 (2020).

\bibitem{Sereshkeh2017online}
Sereshkeh, A. R., Trott, R., Bricout, A., and Chau, T.: Online EEG Classification of Covert Speech for Brain–Computer Interfacing. Int. J. Neural Syst. 27(8), 1750033 (2017).

\bibitem{Lee2019towards}
Lee, S.-H., Lee, M., Jeong, J.-H., and Lee, S.-W.: Towards an EEG-based Intuitive BCI Communication System Using Imagined Speech and Visual Imagery. IEEE Int. Conf. Syst. Man Cybern. (SMC), Italy, pp. 4409--4414 (2019).

\bibitem{Hernadez2021toward}
Hern{\'a}ndez-Del-Toro, T., Reyes-Garc{\'\i}a, C. A., and Villase{\~n}or-Pineda, L.: Toward asynchronous EEG-based BCI: Detecting imagined words segments in continuous EEG signals. Biomed. Signal Process. Control 65, 102351 (2021).

\bibitem{Schirrmeister2017deep}
Schirrmeister, R. T., Springenberg, J. T., Fiederer, L. D. J., Glasstetter, M., Eggensperger, K., Tangermann, M., Hutter, F., Burgard, W., and Ball, T.: Deep learning with convolutional neural networks for EEG decoding and visualization. Hum. Brain Mapp. 3(11), 5391--5420 (2017).

\bibitem{Lawhern2018eegnet}
Lawhern, V. J., Solon, A. J., Waytowich, N. R., Gordon, S. M., Hung, C. P., and Lance, B. J.: EEGNet: a compact convolutional neural network for EEG-based brain-computer interfaces. J. Neural Eng. 15(5), 1--30 (2018).

\bibitem{Schmidhuber2015deep}
Schmidhuber, J.: Deep learning in neural networks: an overview. Neural Netw. 61, 85--117 (2015).

\bibitem{Ronneberger2015unet}
Ronneberger, O., Fischer, P., and Brox, T.: U-net: convolutional networks for biomedical image segmentation. Int. Conf. Med. Image Comput. Comput.-Assisted Intervention (MICCAI), Germany, pp. 234--241 (2015).

\bibitem{Berrar2019cross}
Berrar, D.: Cross-validation. Ency. Bioinform. Comput. Biol. 542--545 (2019).

\bibitem{Ang2008fbcsp}
Ang, K. K., Chin, Z. Y., Zhang, H., and Guan, C.: Filter bank common spatial pattern (FBCSP) in brain-computer interface. IEEE Int. Joint Conf. Neural Netw. (IJCNN), China, pp.2390-–2397 (2008).

\bibitem{Cho2018classification}
Cho, J.-H., Jeong, J.-H., Shim, K.-H., Kim, D.-J., and Lee, S.-W.: Classification of hand motions within EEG signals for non-invasive BCI-based robot hand control. IEEE Int. Conf. Syst. Man Cybern. (SMC), Japan, pp. 515--518 (2018).
\end{thebibliography}
%

\end{document}